\documentclass[11pt]{article}
\usepackage{amsmath,epsfig}
\usepackage{comment,url}

\begin{document}
\baselineskip=18pt

\newcommand{\sn}{\sum_{i=1}^n}
\newcommand{\snj}{\sum_{j=1}^n}
\newcommand{\intt}{\int_0^t}
\newcommand{\inti}{\int_0^{\infty}}
\newcommand{\inttau}{\int_0^{\tau}}

\begin{center}
{\Large \bf Relaxed covariate overlap and margin-based causal effect estimation} \\
\end{center}

\begin{center}
Debashis Ghosh \\
Department of Biostatistics and Informatics, University of Colorado School of Public Health, Aurora, CO 80045\\
\end{center}
\vspace{.001mm}

\begin{center}
{\large \bf \ Abstract}
\end{center}
In most nonrandomized observational studies, differences between treatment groups may arise not only due to the treatment but also because of the effect of confounders. Therefore, causal inference regarding the treatment effect is not as straightforward as in a randomized trial. To adjust for confounding due to measured covariates, a variety of methods based on the potential outcomes framework are used to estimate average treatment effects.  One of the key assumptions is treatment positivity, which states that the probability of treatment is bounded away from zero and one for any possible combination of the confounders.  Methods for performing causal inference when this assumption is violated are relatively limited.    In this article, we discuss a new balance-related condition involving the convex hulls of treatment groups, which I term relaxed covariate overlap.   An advantage of this concept is that it can be linked to a concept from machine learning, termed the margin.  Introduction of relaxed covariate overlap leads to an approach in which one can perform causal inference in a three-step manner.  The methodology is illustrated with two examples.  


\newpage

\section{Introduction}

For many scientific settings, researchers wish to understand the effect of an intervention on a response.   While randomization of the intervention and its evaluation in a prospective clinical trial can provide strong assessments in many instances, for other situations, it is not possible to conduct such a study due to administrative and/or ethical constraints.  Thus, many investigators are left with having to evaluate effects of interventions in observational, non-randomized studies.   This has spawned much research interest in the area of causal inference primarily based on use of the potential outcomes framework (e.g., \cite{imbensrubin}).  

Recently, much attention in the literature on causal inference has been paid to the issue of covariate balance.   This has to do with ensuring that the distributions of confounders between the treatment and control groups have sufficient overlap.   This is related to the treatment positivity assumption that is outlined in \S 2.1.  One way covariate balance is checked in practice is by comparing distributions of individual confounders between the two treatment groups in matched samples (e.g., Chapter 14 of \cite{imbensrubin}).    Matching is typically performed to have robust estimation of a causal effect.   Here, robustness means that estimation of the causal effect does not require extrapolation of the potential outcomes to regions of the covariate space in which observations from one treatment group is missing.  This phenomenon is nicely illustrated in a simple one-dimensional example in Figure 1.   
\begin{figure}[htbp!]\begin{center}
\epsfig{file=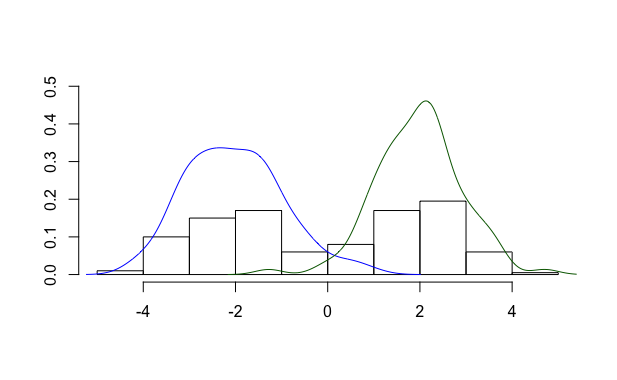,height=2.5in,width=6.5in}
\caption{Histogram of 200 observations, 100 of which are simulated from a normal distribution with mean $-2$ and unit variance, and the other 100 are simulated from a normal distribution with mean $2$ and unit variance.  The blue density corresponds to the former population, while the green density corresponds to the latter.}
\end{center}
\end{figure}
The blue and green density lines represent two different populations.  Towards the left of the figure, most of the observations come from the blue population, while the reverse is true in the right-hand side of the picture.     Causal inference is about attempting to infer differences in the outcomes between the blue and green populations.  The portion of the picture where robust causal inference could be performed would be the region where the densities of the two populations intersect, which is in the middle of the picture.   If one wished to make causal inference in the left part of the picture, this would require model-based extrapolation of outcomes for the green group, and conversely, for the right-hand side of the picture, model-based extrapolation of outcomes for the blue group would be needed.  Thus, I am defining any situation where model-based extrapolation is needed as not being robust.

There has been great attention paid to the use of matching techniques for causal effect estimation \cite{hansen2004, stuart2010, iacus2011}.   A relatively new thread of statistical research has been to focus on estimation procedures that seek to optimize covariate balance in causal effect modelling.  This can be done either by modelling the propensity score to satisfy covariate balance \cite{cbps}, using calibration estimators originally introduced in the survey sampling literature that will satisfy covariate balance \cite{chan2015} or by matching \cite{zubi2015}.  These procedures have been shown to yield weights that are less extreme and lead to causal effect estimators with better properties.   

One situation where covariate balance does not occur is limited treatment overlap, described in \cite{crumpoverlap}.  They characterized its effects in a setting with limited numbers of covariates and developed a simple rule to exclude subjects based on the propensity score.  The procedure of Crump et al. \cite{crumpoverlap} relies on having available propensity score estimates that are consistent for the true propensity score.   There are two limitations with their methodology.  First, the procedure might not be very robust to model misspecification.  Second, it has been been pointed out by several authors that the propensity score might have problems in higher dimensions.   To address this, one proposal was given in \cite{traskinsmall} and consisted of using classification and regression trees (CART).  It takes the Crump et al. \cite{crumpoverlap} definition of a study population with sufficient overlap and then fits a classification tree to whether or not the subject is in the final population or not.  The procedure in \cite{traskinsmall} leads to interpretable regions for which one can define a study population for which one can make causal inferences about.   

The proposals of \cite{crumpoverlap} and \cite{traskinsmall} amount to region identification in which there is sufficient covariate balance between the treatment and control groups.   Similarly, Ratkovic \cite{ratkovictr} developed an approach to causal effect estimation based on support vector machines \cite{svmbook}.  In this article, I focus on the use of the margin for causal effect estimation.    The contributions in this paper are the following:  
\begin{enumerate}
\item Development of a characterization of covariate overlap in a multivariate sense, termed {\it relaxed covariate overlap}, using geometric ideas and relating the problem to margin-based classification.
\item Development of a simple three-step approach to causal inference estimation that avoids the tautology of covariate balance checking.
\item Extension of the margin-based approach to multicategorical and continuous treatments.  
\end{enumerate}
As described in \cite{ghoshlasso}, targeting the observations where covariates overlap should lead to causal effect estimation that does not require model extrapolation and typically runs counter to what standard classifiers wish to do, which is to find maximal separation between populations.   It will be seen that in the multivariate case, relaxed covariate overlap can be characterized using linear hyperplanes and can thus be tied to the support vector machines.  An implication of the methodology is that by excluding subjects, the causal estimand effectively becomes data-adaptive.  We provide some justification for the use of data-adaptive estimands in \S 3.1. and \S 3.2.  The structure of this paper is as follows.   In Section 2, we summarize the potential outcomes framework and describe work in \cite{crumpoverlap} and \cite{traskinsmall} on methods for causal effect estimation in situations where treatment positivity is violated.   Section 3 introduces the relaxed covariate balance condition and demonstrates how it can be tied to a well-known problem in computational geometry, that of determining overlaps of convex hulls of points.   Based on the equivalence, I outline a three-step approach to causal effect estimation.   In Section 4, two examples are used to illustrate the methodology.  Some discussion concludes Section 5.

\section{Background}

\subsection{Preliminaries and Causal inference assumptions}

In this paper, I will employ the potential outcomes framework \cite{neymanpo,rubinpo}, which has been widely used in causal modelling.   I assume the Stable Unit Treatment Value Assumption (SUTVA), which states that the potential outcomes for subject $i$ is statistically independent of the potential outcomes for all subjects $j, j \neq i$, $i,j=1,\ldots,n$.   
Let $Y$ denote the response of interest and ${\bf Z}$ be a $p$-dimensional vector of confounder. Let $T$ be a binary indicator of treatment exposure that takes the values $\{0,1\}$, where $T=1$ if treated and $T=0$ if control.  Let the observed data be represented as $(Y_i,{\bf Z}_i,T_i)$, $i=1,\ldots,n$, a random sample from $(Y,{\bf Z},T)$. Define $\{Y_i(0),Y_i(1)\}$ to be the potential outcomes under control and treatment for subject $i$, where $i=1,\ldots,n$. 

What the analyst observes is $Y_i \equiv Y_i(T_i) = Y_i(1)T_i+ Y_i(0)(1-T_i)$, which implies that $Y_i(0)$ and $Y_i(1)$ can not be observed simultaneously for the $i$th subject.  Two possible parameters of interest are the average causal effect:
\begin{equation}\label{ace}
ACE =E[Y(1) - Y(0)],
\end{equation}
and the average causal effect among the treated:
\begin{equation}\label{acet}
ACET = E[Y(1)-Y(0)|T=1].
\end{equation}
ACET is of particular interest when the population of the study are those who actually receive the treatment. For example, a smoking cessation researcher may wish to know that for those who actually smoke, what is the difference in the expected life expectancy if they did not smoke? In this example, the researcher is interested in estimating ACET.  

Here and in the sequel, I focus on ACE.  
In a randomized study, the treatment assignment is completely determined by randomization. Therefore, $T \perp\{Y(0),Y(1)\}$. Consequently, an unbiased estimator for ACE is given by
\begin{equation*}
\widehat{ACE} =\frac{\sum_{i=1}^n Y_i T_i}{\sum_{i=1}^n T_i}- \frac{\sum_{i=1}^n Y_i(1- T_i)}{\sum_{i=1}^n (1-T_i)}.
\end{equation*}
In an observational study, the vector of covariates ${\bf Z}$ could be related to both the outcome and the treatment assignment. Since both $T$ and the potential outcomes \{Y(0),Y(1)\} are affected by ${\bf Z}$, $ T\perp\{Y(0),Y(1)\}$ will not hold. To enable causal inference in this scenario, I make the following further assumptions.   
\begin{enumerate}
\item Strongly Ignorable Treatment Assumption (SITA):  $\{Y(1),Y(0)\}$ is independent of $T$ given ${\bf Z}$.  
\item Treatment Positivity Assumption (TP):  $1 > P(T = 1|{\bf Z}) > 0$ for all ${\bf Z}$ values.
\end{enumerate}
 SITA means that the potential outcomes are conditionally independent of treatment given the confounders.   
Conceptually, an implication of SITA is that by conditioning on the same value of {\bf Z}, we can assume that the observed outcomes behave as if they came from a randomized study.  Rosenbaum and Rubin \cite{propscore} show that if SITA holds, then the treatment is independent of the potential outcomes given the propensity score, defined as $e({\bf Z}) = P(T = 1|{\bf Z})$.  It is also referred to as the `no unmeasured confounders' assumption in the statistical literature \cite{robins}.  The positivity assumption TP means that the probability of receiving treatment is positive for any individual in the study.    I note that this assumption could be relaxed to the following: $1 > P(T = 1|{\bf Z} = {\bf z}) > 0$ whenever $P({\bf Z} = {\bf z}) > 0$.   

In practice, the TP assumption ensures sufficient covariate overlap.   Balance is necessary in order to develop reliable estimates of causal effects that do not rely on model extrapolation.  There has been a lot of work on developing reliable balance metrics (e.g., \cite{matchit,gmatch}).  However, these procedures implicitly rely on treatment positivity.  I next summarize the proposals of \cite{crumpoverlap} and \cite{traskinsmall} to diagnose and correct for violations in the treatment positivity assumption.    

\subsection{Related work}

Crump et al. \cite{crumpoverlap} noted the possibility that assumption (2), that of treatment positivity, could be violated.   In this case, one conceptual potential outcome of the individual will never be observed.  In  \cite{crumpoverlap}, the authors define a subpopulation covariate effect using the propensity score.  Define the region as ${\cal A}$.    It will be of the form ${\cal A} \equiv \{ {\bf Z}: c \leq e({\bf Z}) \leq 1 - c \}$ for some $c > 0$.   Suppose we observe the data $\{Y_i(1),Y_i(0),{\bf Z}_i\}$, $i=1,\ldots,n$.   Assume also that the propensity score is known.   Then Crump et al. \cite{crumpoverlap} define the subpopulation average causal effect as
 $$ {ACE_{\cal A}} = \frac{\sum_{i=1}^n I\{e({\bf Z}_i) \in {\cal A}\} \{Y_i(1) - Y_i(0)\}}{\sum_{i=1}^n I\{e({\bf Z}_i) \in {\cal A}\}}.$$
Note that ${ACE_{\cal A}}$ depends on the region of the propensity scores that is in ${\cal A}$.   In practice, the propensity score is not known and must be estimated from the data.   

As pointed out in \cite{crumpoverlap}, construction of the region ${\cal A}$ leads to a tradeoff.  The sample size is reduced from $n$ to $nP(e({\bf Z}) \in {\cal A})$, which will lead to increased variability of estimated effects.  On the other hand, the narrowing of the population to subjects whose covariate values are sufficiently balanced will tend to lead to diminished variability in the causal effect estimates.  To simplify calculations, Crump et al. \cite{crumpoverlap} base inference for the subpopulation covariate average effect conditional on the region; in other words, they ignore variability in the estimation of the region ${\cal A}$.   Conditional on the first-stage, they propose an optimization criterion based on the variability of the estimated subpopulation average causal effect.  Crump et al. \cite{crumpoverlap} show under some mild assumptions that an optimal $c^*$ exists, depends only on the marginal distribution of the propensity scores and propose a simple algorithm for its estimation.  The search algorithm is one-dimensional due to the dimension reducing property of the propensity score from $p-$dimensions to a scalar, but it also highlights the dependence on the fitted propensity score model.   If the propensity score model is misspecified, then the approach in \cite{crumpoverlap} might lead to very biased results.   An alternative approach that is more robust was done using classification and regression trees in \cite{traskinsmall}.   Classification and regression trees fit piecewise constant partitions to the observed covariate space.    While the Traskin and Small \cite{traskinsmall} algorithm should be more robust to model misspecification relative to the proposal in \cite{crumpoverlap}, tree models have some limitations as well.  In particular, the types of variables that are fit to the data are axis-parallel planes, which might be too restrictive a class.  In addition, the categorization of observations in \cite{traskinsmall} depend on good estimates of the propensity score.  If there is misspecification in that step, then the tree models are being effectively fit to misspecified outcomes.   Finally, the tree model fit is based on observed covariates.  Thus, the procedure works well if there are covariates that determine the sufficient overlap.  If this is not true, then the proposed procedure will still not provide sufficient balance.  
 
It should also be noted that there are in fact two types of violations of positivity assumptions to consider.  The first is inherent to the model itself, while the second has to do with violations given a finite dataset.   The former might be referred to as a ``structural'' positivity assumption violation, while the latter is a `practical' positivity assumption violation.   Here, we mostly deal with the latter situation.   

\section{Proposed Methodology}

\subsection{Theoretical considerations}

Before describing the proposed approach, it is important to point out that it will effectively amount to defining a data-driven causal estimand.   While there has been a major focus in causal inference to define the appropriate scientific estimand before data collection and to thus avoid the use of data-driven causal estimands, I provide some justification for this approach in the current setting.   

While \S 2.1. discusses the TP assumption in the potential outcomes framework, I recall the work of \cite{rr}, who show that for semiparametric estimators of the average causal effect to have regular behavior in the high-dimensional case, the model classes for the propensity score and outcome models have to be well-behaved.   For the propensity score, this involves strengthening the TP assumption to the following:
\begin{equation}\label{strictoverlap}
\eta < e({\bf Z}) < 1- \eta \ \ \ \ \ \text{w. prob. 1}
\end{equation}
for some $\eta \in (0,0.5)$ uniformly in ${\bf Z}$.  
Note that this means that the propensity score is uniformly bounded away from zero and one.  This is different from the TP assumption in that the latter does not require uniform boundedness away from zero and one.  
Violations in (\ref{strictoverlap}) lead to irregularities in estimation and inference.   It it stronger than the treatment positivity assumption in that $\eta$ does not depend on ${\bf Z}$.   Recently, \cite{damour} show how (\ref{strictoverlap}) implies a bounded likelihood ratio for the distributions of confounders conditional on treatment groups.  The implication is that (\ref{strictoverlap}) becomes a more restrictive assumption as the number of covariates increases.    However, if this assumption does not hold, then this allows for `pathological' data-generating distributions as described in \cite{rr} that lead to causal effect estimators with irregular asymptotic properties.  This issue was described in \cite{gl} and \cite{zlg}.   The approach of identification of the margin as well as estimating  conditioned causal effects, while being data-adaptive, potentially avoids the problem of irregular statistical behavior that would plague average causal effect estimators in the high-dimensional confounder setting.  

\subsection{Practical considerations and analogy with propensity score matching}

The class of methods considered in this paper involves defining data-adaptive causal estimands.   This is similar to the proposals in \cite{crumpoverlap} and \cite{traskinsmall}.   We wish to point out that another popular approach to estimation of causal effects, that of propensity score matching, implicitly involves use of a data-adaptive estimand.   Typically, treated and nontreated subjects with similar propensity scores are matched to each other, and some subjects are excluded from the matched dataset if comparable subjects from the other treatment group cannot be found.  Causal effect estimation then proceeds on the matched dataset.  While this is a commonly used approach to estimation of causal effects, I note that what is being estimated is a causal estimand that is data-adaptive in nature.  

Much of my approach in the current article mimics what is done with propensity score matching.   As a heuristic, analysts who engage in the use of propensity score matching typically take the following steps:
\begin{itemize}
\item[(a).] Fit a propensity score model to the data;
\item[(b).] Match using some algorithm based on the propensity score;
\item[(c).] Check for balance in covariates based on the matched dataset; if there is imbalance, repeat steps (a)-(c). using transformations
of covariates that violate the balance condition.
\item[(d).] Estimate the causal effect in the matched dataset.  
\end{itemize}
For each of the steps there is a variety of choices one can use.   However, as discussed in Theorem 10.1 of  \cite{imbensrubin},  at the last step, there is no adjustment for the standard errors.  They show from their theorem that this variance estimator will overestimate the true variance so that any inferences being made will be conservative.  I adopt the same approach to inference in this paper.  

I note that the approach taken in the paper is but one method with which to handle the issue of limited treatment positivity.   A fuller account can be found in \cite{petersensmmr}, but alternative approaches include restricting the space of treatments, redefining the causal estimand, and using alternative projection functions.  

\subsection{Geometric viewpoint}

To motivate the methodology, I first introduce the concept of a convex hull for a set of points.   

\noindent {\bf Definition.}  Let ${\cal A}$ be a set of points $a_1,\ldots,a_m \in R^d$.  Then the convex hull of ${\cal A}$ is given by 
$$ \text{co}({\cal A}) = \left \{ \sum_{i=1}^m c_i a_i:  c_i \geq 0 \ \ \  \forall i =1,\ldots, m, \sum_{i=1}^m c_i = 1 \right \}.$$
Thus, one sees that the convex hull consists of all convex combinations of points in ${\cal A}$.    There are characterizations of a convex hull equivalent to the definition given here, including the unique minimal convex set containing ${\cal A}$, the intersection of all convex sets containing ${\cal A}$, and the union of all simplices whose vertices are points in
${\cal A}$.  Intuitively, a convex hull will be a multidimensional regions that is `filled up" and also has no `holes' in it.  In addition, by looking at combinations of the observed data points, then by definition any interpolation being done in the generation of the convex hull is a function of the data points only and avoids the model extrapolation issue referred to in the Introduction.   Further details on convex hulls and related topics can be found in \cite{schneider} and \cite{grunbaum}.

A natural extension of covariate overlap to the multivariate case is to require that $\text{co}({\bf Z}|T = 0) = \text{co}({\bf Z}|T = 1)$.  However, finding convex hulls in higher dimensions is a very computationally challenging problem, and Chazelle \cite{chazelle} has shown that the computational complexity of the problem is $O(n^{\lceil{p/2}\rceil})$.  I instead focus on the problem of a nonempty intersection of the convex hulls, i.e.
$$ \text{co}({\bf Z}|T = 0) \bigcap \text{co}({\bf Z}|T = 1)$$
is non-empty.    We refer to this condition as {\it relaxed covariate overlap}.   
 Let the number of subjects with $T = 0$ and $T = 1$ be $n_0$ and $n_1$ respectively.  Denote the $n_0 \times p$ and $n_1 \times p$ matrix of confounders by ${\bf Z}_0$ and ${\bf Z}_1$.  
 
This discussion naturally leads into consideration of the following optimization problem:  minimize over $\alpha$ and $\beta$
\begin{eqnarray}
Q(\alpha,\beta) &=& \frac{1}{2} \| {\bf Z}_0'\alpha - {\bf Z}_1'\beta \|^2 \label{A} \\
s.t.& & \alpha_i \geq 0 \ \ (i = 1,\ldots,n_0) \nonumber  \\
& &  \beta_j \geq 0 \ \  (j=1,\ldots,n_1) \nonumber \\ 
\sum_{i=1}^{n_0} \alpha_i &= 1& = \sum_{j=1}^{n_1} \beta_j \nonumber
\end{eqnarray}
We note that (\ref{A}) is equivalent to the problem of finding the two closest points in the convex hulls $ \text{co}({\bf Z}|T = 0)$ and $ \text{co}({\bf Z}|T = 1)$.  If there is a solution to (\ref{A}), then the convex hulls do not overlap and means that there is no covariate balance between the treated and control groups.  Equivalently, the relaxed covariate overlap condition is not satisfied.  Conversely, if there is no solution to (\ref{A}), then relaxed covariate overlap between the $T = 0$ and $T = 1$ populations is satisfied.   I next derive the dual optimization problem of (\ref{A}).

\noindent {\bf Theorem 1.}  Minimizing (\ref{A}) over ${\bf w}, \alpha$ and $\beta$ is the dual optimization problem of minimizing 
\begin{eqnarray}
\tilde Q(w, \alpha,\beta) =& \frac{1}{2} \| {\bf w}  \|^2 - (\alpha - \beta) \label{B} \\
s.t.& {\bf Z}_0{\bf w} - \alpha {\bf e} \geq {\bf 0} \nonumber  \\
 & -{\bf Z}_1{\bf w} - \beta {\bf e} \geq {\bf 0}, \nonumber  
\end{eqnarray}
over $\alpha$ and $\beta$.  

\noindent {\bf Proof:} The Lagrangian associated with (\ref{B}) is given by 
$$ L \equiv L({\bf w},\alpha,\beta,{\bf u},{\bf v}) = \frac{1}{2} \| {\bf w}  \|^2 - (\alpha - \beta) - {\bf u}'({\bf Z}_0{\bf w} - \alpha {\bf e}) - {\bf v}'( -{\bf Z}_1{\bf w} - \beta {\bf e} )$$
We now seek to maximize $L$ with respect to $\alpha$, $\beta$, ${\bf w}$, ${\bf u}$ and ${\bf v}$.  Differentiation with respect to these parameters and setting them equal to zero yields 
\begin{eqnarray}
{\bf w} - {\bf Z}_0'{\bf u} + {\bf Z}_1'{\bf v} &=& {\bf 0} \label{lagrangian} \\
-1 + {\bf e}'{\bf u} &=& 0 \nonumber \\
1 - {\bf e}'{\bf v} &=& 0 \nonumber
\end{eqnarray}
subject to ${\bf u} \geq {\bf 0}$ and ${\bf v} \geq {\bf 0}$.  Plugging in  ${\bf w} = {\bf Z}_0'{\bf u} - {\bf Z}_1'{\bf v}$ and simplifying (\ref{lagrangian}) yields the result of the theorem.

There is a natural interpretation of Theorem 1 as well.  The functions ${\bf Z}_0{\bf w}$ and ${\bf Z}_1{\bf w}$ define hyperplanes, and an implication of Theorem 1 is that there is a solution of (\ref{A}) if and only if there is a solution to (\ref{B}).   The equations ${\bf Z}_0{\bf w} \geq \alpha {\bf e}_0$ and ${\bf Z}_1{\bf w} \leq -\beta{\bf e}_1$ mean that there exists a hyperplane that perfect separates the confounders for the $T = 0$ and $T = 1$ populations.   In machine learning terminology, this scenario corresponds to the observations being linearly separable.   Furthermore, provide there exists a solution to the problem in Theorem 1, then ${\bf Z}'{\bf w} = (\alpha + \beta)/2$ defines the hyperplane that maximizes the distance between the supporting hyperplanes for $\text{co}({\bf Z}|T = 0)$ and $ \text{co}({\bf Z}|T = 1)$.   

The quantity $2/\|{\bf w}\|^2$ is referred to as the margin in the machine learning literature \cite{vapnik}.   The optimization problems (\ref{A}) and (\ref{B}), in words, represent the following equivalence:

\begin{equation*}
\begin{pmatrix}
\text{No overlap in }  \text{co}({\bf Z}|T = 0) \\
\text{and }  \text{co}({\bf Z}|T = 1) 
\end{pmatrix}
\Leftrightarrow
\begin{pmatrix}
\text{Existence of a linear hyperplane that} \\
\text{separates } {\bf Z}|T = 0 \text{ and  }{\bf Z}|T = 1.
\end{pmatrix}
\end{equation*}
If the data are linearly inseparable, then this means that the convex hulls of ${\bf Z}|T = 0$ and ${\bf Z}|T = 1$ will overlap.   The points in the overlap will be identical to those that fall in the margin and represent those observations in which the possibility of causal inference without model extrapolation is feasible.   This is very intuitive in the sense that for points in the margin, it is difficult to classify them into treatment groups, so these are the observations for which the TP assumption will be valid.    This also suggests that to identify observations that satisfy covariate balance, it is important to target the margin as the criterion on which to optimize.  At a high level, the approach being proposed here amounts to the following:

\begin{enumerate}
\item Fit a model to the data $(T_i,{\bf Z}_i)$, $i=1,\ldots,n$.
\item Determine the observations that are in the margin.  Let this set of observations be denoted as ${\cal M}$.  
\item Estimate the causal effect of interest using $(Y_i,T_i, {\bf Z}_i)$, $i \in {\cal M}$.   
\end{enumerate}
Going through this progression, the goal of the first two steps is to identify the observations which are likely to satisfy the treatment positivity assumption.   The first step is typically done using propensity score models, although other models could be entertained at that step.    In step two of the procedure, observations that are not in the margin will be discarded from the analysis.   Thus, the determination of the margin step leads to selection of observations for performing causal inference.   Another approach by which observations get discarded is in matching, where unmatchable observations are removed (e.g., \cite{matchit}).   However, there is no underlying concept of a margin in that approach.  Finally, an advantage of this procedure is that it does not require balance checking, as is the norm in causal inference problems.  

\noindent {\bf Remark 1:} Ratkovic \cite{ratkovictr} developed an algorithm for achieving balance in causal inference problems that uses support vector machines.   Ratkovic identifies the points in the margin as being the relevant ones for causal inference.  However, the derivation presented there involves first-order conditions based on optimizing the penalized loss function corresponding to SVM.    By contrast, I have started from geometric principles of overlapping convex hulls in order to identify the margin.   

\subsection{Support Vector Machines}

As alluded to earlier, support vector machines (SVMs) represent another class of algorithms that seek to optimize the margin.   The objective of SVM is to find a linear hyperplane that maximizes the margin between the populations defined by $T = 1$ and $T = -1$.   SVMs are formulated using the following optimization problem: minimize as a function of ${\bf w}$ and $b$ the norm of ${\bf w}$ 
subject to $T_i({\bf w}\cdot {\bf Z}_i - b) \geq 1$, $i=1,\ldots,n$.   Here, ${\bf a} \cdot {\bf b}$ denotes the inner product between vectors ${\bf a}$ and ${\bf b}$, with $\|{\bf a}\| = ({\bf a} \cdot {\bf a})^2$.    Note that $1/\|{\bf w}\|$ is proportional to the margin.   Finding the hyperplane that maximizes the margin is equivalent to minimizing the square of the inverse of the margin.  This turns out to be a quadratic programming problem and can be phrased formally using Lagrange multipliers as 
\begin{equation}\label{svmopt}
\arg \min_{{\bf w},b} \max_{\alpha \geq 0} \left \{ \frac{\|{\bf w}\|^2}{2} - \sum_{i=1}^n \alpha_i [T_i({\bf w}\cdot {\bf Z}_i - b) - 1] \right \}
\end{equation}
Using the Karuhn-Kush-Tucker conditions from optimization theory, it turns out that the solution can be represented as $ {\bf w} = \sum_{i=1} \alpha_i T_i {\bf Z}_i$, where only a subset of the $n$ observations will have $\alpha_i > 0$.  The remaining observations will have $\alpha_i = 0$.   The subjects for which $\alpha_i > 0$ $(i=1,\ldots,n)$ are termed the {\it support vectors}.   It turns out the margin depends only on the support vectors.   Thus, one of the appealing features of SVMs is that they are sparse in the observations in the dataset.   This yields a very simple classification rule.   If ${\bf w}\cdot {\bf Z}_i - b \geq 1$, then predict $T = 1$; if ${\bf w}\cdot {\bf Z}_i - b \leq -1$, predict $T = -1$.    Thus, using the arguments in \S 3.1., if the data are linearly separable using this hyperplane, then one would not have relaxed covariate overlap.  Thus, the observations that violate the classification rule, which are equivalent to the misclassified observations, represent the margin for which I will use to perform causal inference.   Intuitively, this makes sense as points in the margin represent those points about which there is uncertainty as to the classification of the subject (i.e., $T = 1$ or $-1$), while for those points outside of the margin,  one is quite certain as to their treatment label.   The key point is that the SVM defines the region ${\cal M}$, which represents the part of the sample for which a causal effect will be estimated.   This is in keeping with the idea posited in \cite{rosenbaumdesign} that the population under study can be a subpopulation of the original population based on observed covariate values.  

To graphically illustrate the concept, I simulated data using two bivariate normal populations.  The plotted data and the fitted SVM-based margin are shown in Figure 2.  
\begin{figure}[htbp!]\begin{center}
\epsfig{file=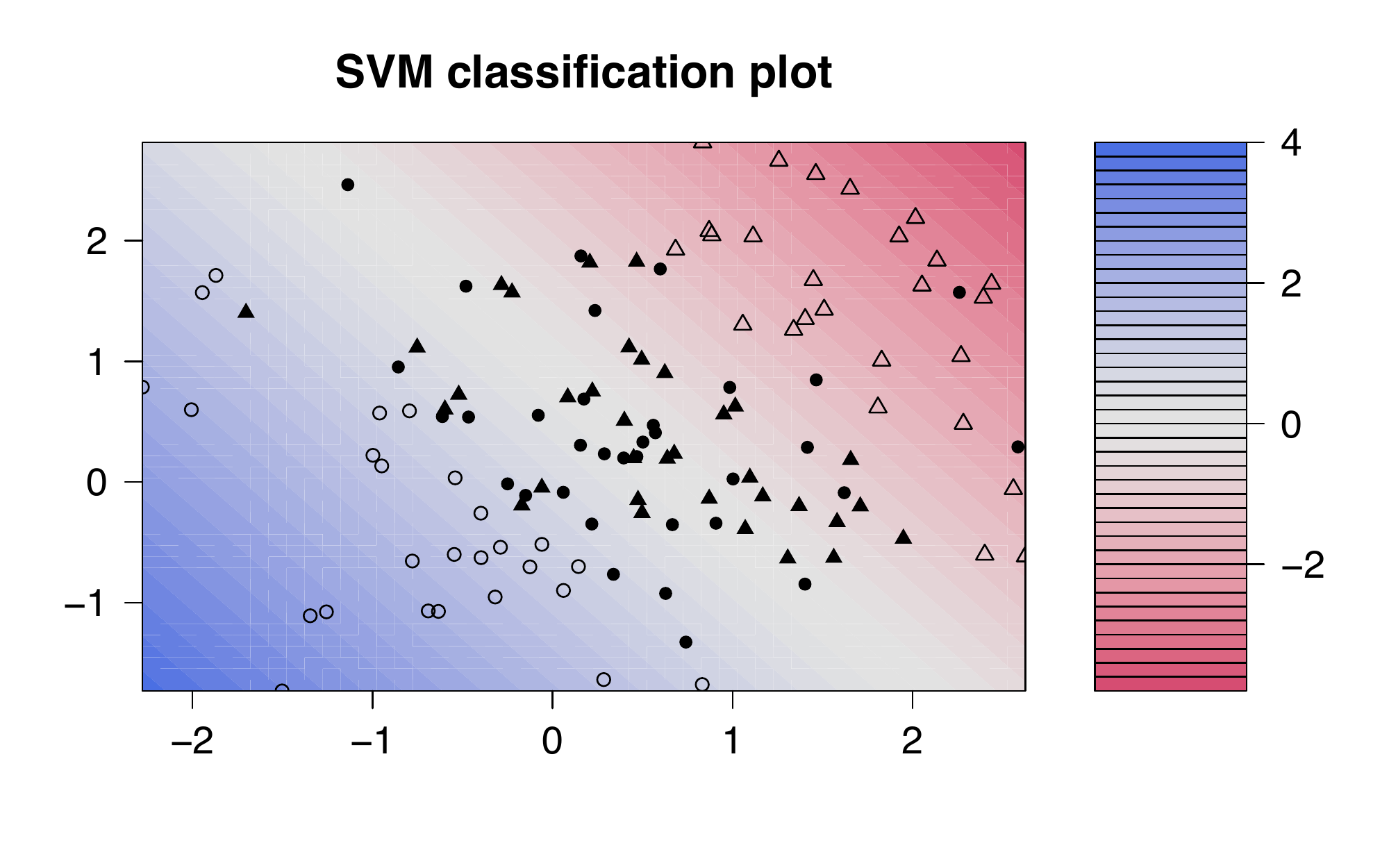,height=4in,width=6.5in}
\caption{Example of the margin concept in a bivariate normal example.  I have simulated two populations, represented as circles and triangles.   The circles are generated from a bivariate normal distribution with mean $(0,0)$, while the triangles are generated from a bivariate normal distribution with mean $(1,1)$.  Both distributions have an identity variance-covariance matrix.  I have fit a support vector machine to the data using a linear kernel with cost parameter $C = 1$.  The observations that are in the margin are filled in (i.e., the filled-in circles and triangles).  The proposal is to use the black points for causal effect estimation.}
\end{center}
\end{figure}

\subsection{Causal effect estimation, inference and other outputs}
In this article, I focus on the estimation of a particular subpopulation average causal effect.  Recall the general three-step approach outlined in \S 3.3. :  (a) fit a propensity score model to the data $(T_i,{\bf Z}_i)$, $i=1,\ldots,n$; (b) identify the observations in the margin based on the model fit in (a) and label the observations in the margin as ${\cal M}$; (c) perform causal effect estimation using  $(Y_i,T_i, {\bf Z}_i)$, $i \in {\cal M}$.    The goals of steps (a) and (b) are to define the subdataset on which I will estimate a causal effect.   This is keeping with the principle outlined in \cite{matchit} that there can be a preprocessing step in which observations are discarded before beginning to perform causal inference.   The preprocessing here is to remove the observations that violate treatment positivity, which are the observations that do not fall into the margin.    This is also keeping in line with the spirit of \cite{rubin2007}, who advocates separating the outcome model step from the propensity score model step and to not have feedback between two.   Steps (a) and (b) only involve the propensity score model, and an appropriate population of interest gets determined at this stage.   Thus, the inference in step (c) will be conditional in that the margin needs to be found first, and then causal effect estimation in step (c) happens conditional on the margin.   

For step (c), there are many choices available for causal effect estimation.  For the purposes of illustration, here I will use the optimal matching approach described in \cite{hansen2004}. However, as discussed in \cite{lunceforddavidian}, there are many ways to perform causal inference in step (c).   I take the approach of \cite{matchit} in that no adjustment needs to be made to the standard errors based on the analysis in the matched sample.   Evaluating various approaches to standard error estimation for the causal effect in the matched sample setting remains an open topic and the subject of future investigation.       

One of the features of the proposed use of the margin for causal inference is that one can assess violations in the TP assumption.   Recall that this means that $1> P(T = 1|{\bf Z}) > 0$ for all values of ${\bf Z}$. In words, this means that the probability of receiving treatment is positive for any individual in the study. However, there may be many situations when this assumption will be violated. This could occur for practical or empirical reasons. In the idealized setting of a clinical study, only certain individuals might be allowed to receive treatment based on observed covariate values.  Points that are outside of the margin will represent those where TP is violated.   As in \cite{traskinsmall}, I could model being in the margin using a classification and regression tree, which might provide a nice interpretable descriptive summary of what factors define being in the margin.  

\noindent {\bf Remark 2:} There has been much success in combining the propensity score modelling with the mean outcome modelling for causal modelling using collaborative targeted learning \cite{gl}.  In fact, in Chapter 21 of \cite{vdr}, it is shown that this approach can successfully deal with problematic causal inference scenarios.     In principle, it may be possible to extend the targeted learning roadmap to include margin identification as part of the approach, but this is beyond the scope of the current manuscript.

\subsection{Extension to multicategorical treatments}

One of the advantages of the geometric notion of the overlap described in \S 3.3. is that it admits a natural extension of multicategorical treatment variables and thus allows for a natural extension of the causal inference approach to multiple treatments.   I note that the problem has received less attention than the binary case, although some exceptions include \cite{imbensdose,gps}.  

I need to modify the assumptions from \S 2.1. to accomodate multiple treatment levels.  Let $T^*$ take values $\{0,1,\ldots,K-1\}$.  I then make assumptions generalizing those in \S 2.1:
\begin{enumerate}
\item The potential outcomes for subject $i$ is statistically independent of the potential outcomes for all subjects $j, j \neq i$, $i,j=1,\ldots,n$.   
\item  $\{Y(0),\ldots,Y(K-1)\}$ is independent of $T$ given ${\bf Z}$.  
\item $P(T = k|{\bf Z}) > 0$ for all ${\bf Z}$ values and for all $k=0,\ldots,K-1$.   
\end{enumerate}
Arguing as in \S 3.3., one can get to the following equivalence:
\begin{equation*}
\begin{pmatrix}
\text{No overlap in }  \text{co}({\bf Z}|T = i) \\
\text{and }  \text{co}({\bf Z}|T = j) \\
\text{for any $i,j$}
\end{pmatrix}
\Leftrightarrow
\begin{pmatrix}
\text{Existence of a hyperplane that} \\
\text{separates } {\bf Z}|T = i \text{ and  }{\bf Z}|T = j\\
\text{for any $i,j$.}
\end{pmatrix}
\end{equation*}
Thus, one could envision performing either a multi-class SVM in order to derive the margins.    The approach in the multicategorical case is to take the union of the margins from all $\begin{pmatrix} K \\ 2 \end{pmatrix}$ classifiers as the meta-margin and to perform appropriate causal comparisons based on pairwise treatment comparisons.  This is described in the choliangocarcinoma example in \S 4.1. 

\subsection{Extension to continuous treatments}

As described in \cite{gps} and \cite{zhujci}, there can be situations in which the causal estimand of interest is defined based on a variable that is continuous.   For these situations, development of the margin is not intuitive at first glance.    I will use the arguments in \cite{bibennett} in order to extend the margin-based causal approach to accommodate continuous treament variables. 

In the previous sections, I have fit support vector machines for treatment.  In the case where it is continuous, a natural analog is support vector regression.   Statistically, this can be be expressed as the following optimization problem: for a fixed $\epsilon > 0$, 
$$ \min_f \sum_{i=1}^n |T_i - f({\bf Z}_i)|_{\epsilon} + \lambda \|f\|^2_{H_K}$$
where $$|u|_{\epsilon} = \max(0,|u|-\epsilon),$$
$\lambda > 0$ is a smoothing parameter and $\|\cdot\|^2_{H_K}$ denotes the norm of a function in a reproducing kernel Hilbert space (RKHS; \cite{wahbarkhs}).  As in \S 3.1., I will choose the RKHS that corresponds to the linear kernel.   I will also define a hard $\epsilon-$tube as a hyperplane such that for $i = 1,\ldots,n$, 

$$  -\epsilon \leq T_i - w'{\bf Z}_i - b \leq \epsilon.$$
It is easy to see that if by choosing $\epsilon > 0$ to be sufficiently large, then a hard $\epsilon-$tube will exist.   Using Gale's Theorem \cite{mangasarian}, a hard $\epsilon-$tube exists if and only if the following system of equations in $({\bf u},{\bf v})$ has no solution:   
$$ ({\bf y} + \epsilon{\bf e})'{\bf u} -  ({\bf y} - \epsilon{\bf e})'{\bf v} < 0$$
$$ {\bf Z}_0'{\bf u} = {\bf Z}_1'{\bf v}, \ \ {\bf e}'{\bf u} = {\bf e}'{\bf v} = 1, \ \ \ {\bf u} \geq {\bf 0}, {\bf v} \geq {\bf 0}.$$

I now define the sets $D^+_{\epsilon} = \{({\bf Z}_i,T_i + \epsilon),i=1,\ldots,n\}$ and $D^-_{\epsilon} = \{({\bf Z}_i,T_i - \epsilon),i=1,\ldots,n\}$.  Using the arguments in \cite{bibennett}, it can be seen that the existence of a hard $\epsilon$-tube is equivalent to the convex hulls of $D^+_{\epsilon}$ and $D^-_{\epsilon}$ being separable.   Thus, I have recast the problem of margin for the support vector regression into the setup presented in \S 3.1.  

Unlike in the binary and categorical data cases, one is unable to use full matching in order to perform causal effect estimation.   Instead, I will adopt the approach used in \cite{zhujci}, which is to use generalized boosting \cite{ridgeway1999} in conjunction with normal model fitting and weighted estimation in order to estimate causal parameters.   Further details about the implementation can be found in the example in \S 4.2.   

\noindent {\bf Remark 3:}  Recently, \cite{robplatt} showed that violations of the treatment positivity assumption in marginal structural models can manifest in observations with very large weights.  In an analysis of a database designed to study the effect of warfarin on the risk of bleeding, they found that the uncritical use of marginal structural models yielded an odds ratio of 17.2, while using restricted weights in the marginal structural models yielded an odds ratio of 2.0.   The latter was much more in line with results in the field.   
\section{Numerical Examples}

\subsection{Choliangocarcinoma Data}

The first example comes from a dataset of 3894 patients with intrahepatic cholangiocarcinomas (IHC) that was previously studied in \cite{choliangocarcinoma}.   In this study, the effect of radiation and surgery on patient was survival in this population was explored using data extracted from the Surveillance, Epidemiology and End Results (SEER) registry.    Note that there are four levels of treatment: no treatment, radiation only, surgery only and combined (radiation and surgery).   While Figure 2 shows some overlap in the plots of overall survival, a log-rank test reveals a highly significant difference between the four groups. 
\begin{figure}[htbp!]\begin{center}
\epsfig{file=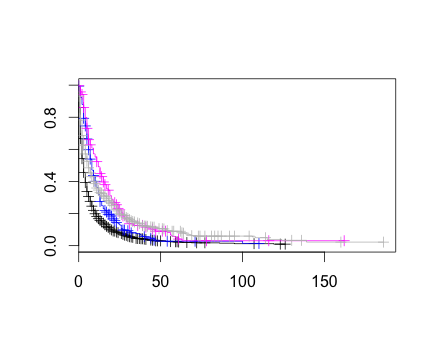,height=3in,width=6.5in}
\caption{Kaplan-Meier curves of overall survival by the treatment group for the IHC study.   The treatment group corresponds to no treatment (black line), radiation only (blue line), surgery only (purple line) and combined treatment (gray line).  The log-rank statistic for comparing the four groups is 258 and is distributed as chi-squared with three degrees of freedom under the null hypothesis of no difference in survival between the groups.}
\end{center}
\end{figure}
Because patients in the SEER registry are not randomized to treatment, there might be self-selection in patients' choice of treatments, leading to confounding.   I illustrate the methods in the paper by first comparing the combined treatment group to the rest.  A proportional hazards model of overall survival on this binary treatment yields an estimated hazard ratio of 0.60 with an associated 95\% confidence inteval of (0.52,.68).  Thus, use of the combined treatment is associated with a 40\% reduction in relative risk of death.   I use the following variables as confounders in the analysis: age, stage of cancer, race and SEER registry location.  If I adjust for these variables in the proportional hazards model, then the estimated hazard ratio of treatment changes to 0.62 with an associated 95\% confidence inteval of (0.55,0.72). 

The analyses in the previous paragraph used all 3894 observations.  I now apply the margin methodology.   To reiterate, this corresponds to the following three steps: (a) fit a support vector machine with combined treatment versus the rest as the outcome; (b) identify the margin observations and perform full matching; (c) fit a proportional hazards regression model of survival on treatment where the matched observations are treated as fixed strata, i.e. each matched set has a separate baseline hazard function in the Cox model, but the covariate effects are the same across matched sets.  I used the \textbf{svm} function available in the \textbf{e1071} library for step (a); we assumed default parameter settings.   This analysis yielded an estimated hazard ratio of 0.42 with an associated 95\% CI of (0.32,0.54).   Thus, this approach leads to a larger effect size.   What is key to note, however, is that we are no longer using all 3894 observations.   This analysis uses 604 observations, so we have in effect discarded over 85\% of the observations.   While much data have been removed, the tradeoff is that the remaining observations better satisfy the treatment positivity assumption.     A simulation exercise was performed where I explored the bootstrap distribution of the number of observations in the margin; the results are shown in Figure 3.   What is seen here is that the modelling for the true effect relies on only 10-20\% of the data.  This underscores the fact that while more observations might be desirable from a statistical point of view, for causal effect estimation problems, a potentially key concept is the margin size.  

\begin{figure}[htbp!]\begin{center}
\epsfig{file=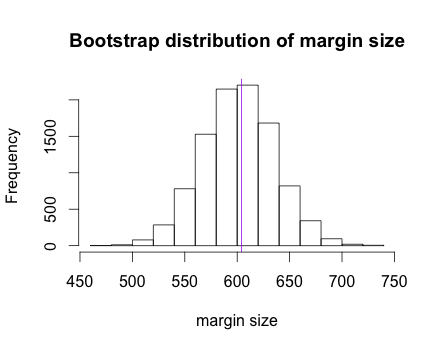,height=2.5in,width=4.5in}
\caption{Bootstrap distribution for the margin size based on the IHC data.   The purple line denotes the margin size for the observed data (604 observations); the distribution is roughly centered around the observed value.  }
\end{center}
\end{figure}

Next, I consider treatment as a four-level variable based on the four groups in Figure 1.   If one fits a multicategorical SVM using all 6 pairs of classifiers, I find that the meta-margin only excludes observations from the no treatment group.   Out of the 2333 IHC subjects who receive no treatment in the original dataset, 1188 are removed.  I then ran analyses with all six possible pairwise treatment comparisons using a regression adjustment strategy versus the proposed methodology.  They are summarized in Table 1.   Based on the results, several findings obtain.   First, there is no reduction in the number of observations for comparisons that do not involve the no treatment group.  For those comparisons, the standard and proposed approaches give fairly concordant results.   More pronounced differences are found in comparisons involving the no treatment group.    In general, the proposed method leads to stronger effect estimates, although this comes as the cost of slightly increased variability.   Combining radiation and surgery tends to lead to decreased risk of death relative to using either treament by itself.   

\begin{table}[htbp!]
\centering
\begin{tabular}{lcccc}
& \multicolumn{2}{c}{Standard} & \multicolumn{2}{c}{Proposed}\\
\hline
Comparison&HR (95\% CI)& n &  HR (95\% CI) & n \\
\hline
Radiation/No Treatment&0.63 (0.55,0.71)& 2676 & 0.49 (0.41,0.59)&1488\\
Surgery/No Treatment & 0.66 (0.60,0.72) & 3289 & 0.64 (0.56, 0.73)& 2101  \\
Combined/No Treatment &  0.51 (0.44, 0.59) & 2595 &  0.37  (0.30, 0.46) & 1407\\
Surgery/Radiation & 1.00  (0.87,1.15) & 1299 &  1.06 (0.88,1.26) & 1299\\
Combined/Radiation & 0.76 (0.65,0.87) & 1218 & 0.60 (0.49,0.74) & 1218 \\
Combined/Surgery & 0.68 (0.57,0.83) & 605 & 0.70 (0.55,0.90) & 605 \\
\hline
\end{tabular}
\caption{Hazard ratios for pairwise treatment comparisons in IHC data using a regression-based PH model (Standard) as well as the margin-based approach (Proposed). HR represents hazard ratio for death, while 95\% CI denotes the associated 95\% confidence interval.  For the comparison column, A/B represents a comparison between groups A and B with B denoting the reference group.  For example, the `Radiation/No Treatment' entry denotes comparing the radiation-treated group to the no treatment group, with the latter serving as the reference group.}
\end{table}
As a final exploratory analysis, I created a variable based on being included in the margin and fit a classification tree model in order to determine what factors determined being in the margin.  The first split is done on age, and if age is less than 63.5 years, then there is a 93\% chance of being in the margin.  This suggests that margin-based inferences are being made with respect to a younger population than what is represented in the entire IHC dataset.   

\subsection{Early Dieting in Girls Study}
To illustrate the margin-based approach with a continuous treatment, we use data from the Early Dieting in Girls study, a longitudinal study in which mother-daughter dyads were followed at five time points.  The study population comes from white non-Hispanic families living in central Pennsylvania.   At each time point, measurements were taken, and the mothers and daughters were also interviewed.  Broadly speaking, the goals of the study are to examine parental influences on daughters' growth and development from ages 5 to 15; further details can be found in 
\cite{edgs}, \cite{zhujci} and \cite{zhujrssc}.

This analysis models the influence of mothers' weight concern in year 2 of the study on their daughter's body mass index at year 3 of the study.   The treatment variable is mother's overall weight concern which is measured when age 7. It is the average score of  five questions in the questionnaire. A higher value implies the mother is more concerned about gaining weight. In the dataset, and its values range from 0 to 3.4.   There were 21 potential baseline confounders considered in this study regarding participants' characteristics, such as family history of diabetes and obesity, family income, daughter's disinhibition, daughter's body esteem, mother's perception of mother's current size and mother's satisfaction with daughter's current body.   The margin-based approach will involve using support vector regression.   I follow the same three-step procedure as described above.   One issue that arises is that for the third step, one cannot use full matching as in the previous example.   Instead, I adopt the approach from Zhu et al. \cite{zhujci} for performing causal inference with a continuous treatment:  
\begin{enumerate}
\item[1.]  Fit $T_i$ on ${\bf Z}_i$ using generalized boosting \cite{ridgeway1999} and for $i=1,\dots,n$, and get $\hat{T}_i$ and $\hat{\sigma}$;
\item[2.]  Calculate the residuals $\hat{\epsilon_i}=T_i-\hat{T}_i$; $r(T_i, {\bf Z}_i)$ can be approximated by
\begin{equation}\label{normal}
\hat{r}(T_i, {\bf Z}_i)\approx f(\hat{\epsilon_i})\approx\frac{1}{\sqrt{2\pi}\hat{\sigma}}\exp\{-\frac{\hat{\epsilon_i}^2}{2\hat{\sigma}^2}\}.
\end{equation}
\item[3.] Compute stabilized weights for $i=1,\ldots,n$,
$$w_i = \frac{\hat r_0(e_i)}{\hat r(T_i,{\bf Z}_i)},$$
where $\hat r_0(e_i)$ denotes the estimated residual for observation $i$ using the null model (i.e., not involving any covariates ${\bf Z}$).  
\item[4]. Run a regression of $Y$ on $T$ using the stabilized weights.
\end{enumerate}
Applying this approach to the full dataset ($n = 159$ observations) yields a causal effect estimate of $0.82$ with a standard error of $0.41$.  This corresponds to a test statistic of  $1.99$, which is marginally significant at the 0.05 level of significance.  Using the margin-based approach with $\epsilon = 0.1$ discards 21 observations, but the effect estimate changes to 0.41 with an associated standard error of 0.76, which is a non-signficant effect.  Thus, there is a large number of observations that violate the relaxed covariate overlap condition and thus might also violate the treatment positivity condition.   

\section{Discussion}

In this article, I have shown how the margin concept from machine learning provides a basis for estimating causal effects in a manner not requiring model extrapolation and leads to a natural three-step approach for causal inference.    The margin identifies regions in the covariate space where this is overlap in the confounders between treatment groups.   Areas where there is no covariate overlap violate key assumptions in causal inference, such as the treatment positivity assumption.   

While the margin from support vector machines has been espoused in \cite{ratkovictr}, there are several important differences between that work between what is proposed here.   The use of SVMs naturally arise from consideration of the duality between covariate overlap with separating hyperplanes.  Further, we only consider linear separating hyperplanes or equivalently, SVMs with a reproducing kernel Hilbert space corresponding to a linear kernel.     By contrast, Ratkovic \cite{ratkovictr} proposes a hierarchical SVM model and attendant Bayesian inferential procedures for linear and nonlinear SVMs.   There is no intuitive geometric covariate overlap notion for what Ratkovic \cite{ratkovictr} proposes in the nonlinear case, and he uses first-order conditions to argue for covariate balance in the margin.   However, that work as well as the current paper argues for better understanding of the statistical properties of the margin.   This is currently under investigation.   

Implicit in the causal effect analyses is that inferences are done conditionally on finding the margin.  As argued in other contexts (e.g., \cite{mode}), there exist many modes of performing inference in causal analyses.  This issue should be explored further as well.   In particular, the recent literature on post-model selection inference in \cite{jtaylor1} could potentially be extended to this setting.  

\section*{Acknowledgement}
This research is supported by a pilot grant from the Data Science to Patient Value (D2V) initiative from the University of Colorado.  The author would like to thank Dr. Yeying Zhu and Dr. Nandita Mitra for providing the dieting and choliangocarcinoma datasets, respectively.  The author would like to acknowledge an associate editor and referee, whose comments greatly improved the quality of the manuscript.

\end{document}